\documentclass[a4paper]{jpconf}
\usepackage{graphicx}

\renewcommand{\thefootnote}{\fnsymbol{footnote}}
\newcommand{\beq}{\begin{equation}}
\newcommand{\eeq}{\end{equation}}

\begin{document}
\title{Cold Dark Matters in $U(1)'$ Models}

\author{Yuji Omura}

\address{School of Physics, KIAS, Seoul 130-722, Korea}

\ead{omura@kias.re.kr}

\begin{abstract}
We consider models with extra $U(1)'$ gauge symmetry that is broken spontaneously.
In the models, there are cold dark matter candidates which are charged under $U(1)'$ symmetry. 
Depending on the charge assignment, we can evade the strong bound from the Drell-Yang process,
and consider the scenario that the dark matters strongly interact with the fermions in the Standard Model
through the extra gauge boson exchanging. As an illustrative example, we discuss a supersymmetric gauged $U(1)_B \times U(1)_L$ model
and see that not only correct relic density but also the DAMA/CoGeNT excess can be achieved by the dark matters.   
\end{abstract}

\section{Introduction}

Many astrophysical observations support the existence of cold dark matter (CDM) in our universe,
but its properties, such as its mass, spin, and interactions, are still largely unknown.
Several decades of experimental work have been spent trying to detect dark matter particles coming to Earth (direct detection), or more generally products of dark matter reactions in outer space (indirect detection).
In the both cases, very interesting signals, which might be originated by dark matters, were observed, but we could have not yet concluded that they are really the evidences of dark matters.  

On the other hand, it is well known that the Standard Model (SM) describes the particle physics, up to the weak scale. Except for the Higgs sector, the SM has been investigated very well and we know that it can realize most of all the experimental results at high accuracy. 
However, one of the insufficient points of the SM might be dark matter physics. 
The SM does not include a heavy neutral stable particle which could be a CDM candidate, 
so that the extension of the SM may be required.

In this letter, we propose the extension to gauged $U(1)'$ symmetric models, motivated by CDM. 
In our models, the additional gauged symmetry is assigned to the SM fermions and extra chiral fermions are also added to satisfy the anomaly conditions. Some of the extra fermions become neutral and stable after the electro-weak symmetry breaking (EWSB). Furthermore, an extra scalar which is only charged under $U(1)'$ is
required to avoid stable extra quarks. It also becomes stable because of the remnant global symmetry after $U(1)'$ symmetry breaking, and interacts with the SM fermions through the $U(1)'$ gauge boson exchanging. When we consider the extension of our models to the supersymmetric, we could realize another CDM candidates such as the superpartner of the additional scalar and neutralino. 

This letter is devoted as follows. In the next section, we introduce our $U(1)'$ models, where the SM particles are charged under $U(1)'$ symmetry and extra fermions are added
for the anomaly conditions. We also see that some of the extra particles become neutral and good CDM candidates in the section 3. The CDMs interact with the SM particles through the new gauge boson exchanging, and could achieve the correct relic density and predict the signal of direct detections. As an illustrative model, we consider a supersymmetric gauged $U(1)_B \times U(1)_L$ model~\cite{Ko-Omura} in the section 4, and estimate the thermal relic density and the cross section of direct detection with proton.

\section{Setups of our models}
\label{sec:anomaly}
We start with $U(1)'$ gauge models with the 
following flavor-dependent charge assignments: 
\begin{center}
\begin{tabular}{ccccccc}\br
          & $SU(3)_c$ & $SU(2)_L$ & $U(1)_Y$ & $U(1)'$    \\ \mr
$Q_i$     & $3$       & $2$       & $1/6$    & $q_{i}$    \\ \hline 
$D_{Ri}$  & $3$       & $1$       & $-1/3$   & $d_i$      \\ \hline
$U_{Ri}$  & $3$       & $1$       & $2/3$    & $u_i$      \\ \hline  
$L_i$     & $1$       & $2$       & $-1/2$   & $l_i$        \\ \hline 
$E_{Ri}$  & $1$       & $1$       & $-1$     & $e_i$        \\ \hline
$H$       & $1$       & $2$       & $1/2$    & $q_h$      \\ \hline 
$\Phi$    & $1$       & $1$       & $0$      & $q_{\Phi}$ \\ \br  
\end{tabular}
\end{center}
with $Q_i^T =(U_{Li},\,\! D_{Li})$, 
$L_i^T =(\nu_{Li},\,\! E_{Li})$  and $i=1,2,3$ are generation indices.  
$\Phi$ is a SM-gauge singlet with nonzero $U(1)'$ charge which is 
required to break $U(1)'$ spontaneously and 
generate nonzero masses for $Z'$ and extra fermions.

The flavor-dependent $U(1)'$ models generally become 
anomalous without extra chiral fields.
One of the simplest ways to construct anomaly-free theory is to add 
one extra generation and two SM gauge vector-like pairs as follows:

\begin{center}
\begin{tabular}{ccccccc}\br
         & $SU(3)_c$ & $SU(2)_L$ & $U(1)_Y$ & $U(1)'$                   \\ 
\mr  
$Q'_R$     & $3$       & $2$       & $1/6$    & $(q_{1}+q_{2}+q_{3})$ \\
\hline 
$D'_L$   & $3$       & $1$       & $-1/3$   & $(d_1+d_2+d_3)$          \\ 
\hline 
$U'_L$   & $3$       & $1$       & $2/3$    & $(u_1+u_2+u_3)$\\ \hline 
$L'_R$     & $1$       & $2$       & $-1/2$   & $(l_1+l_2+l_3)$             \\ \hline 
$E'_L$     & $1$       & $1$       & $-1$     & $(e_1+e_2+e_3)$             \\ \hline  
$l_{L1}$ & $1$       & $2$       & $-1/2$   & $Q_L$           \\ \hline 
$l_{R1}$ & $1$       & $2$       & $-1/2$   & $Q_R$           \\ \hline 
$l_{L2}$ & $1$       & $2$       & $-1/2$   & $-Q_L$          \\ \hline 
$l_{R2}$ & $1$       & $2$       & $-1/2$   & $-Q_R$          \\ \br 
\end{tabular}
\end{center}
 $Q'_R$, $D'_L$, $U'_L$, $L'_R$, and $E'_L$ are like the extra SM fermions of the fourth generation and
 their chiralities are opposite. The $U(1)'$ charges are defined to satisfy the conditions for $U(1)'$, $U(1)'SU(2)^2_L$, $U(1)'SU(3)^2_c$ and $U(1)'U(1)_Y^2$. We just assume that they also satisfy the condition for $U(1)'^3$, but
there might be not so many simple solutions, if one requires rational numbers for all the $U(1)'$ charges.  
$Q_{L,R}$ contribute to only $U(1)_YU(1)'^2$, and we can easily find the solution with $Q_L-Q_R=1$, as discussed in  Ref.~\cite{Ko1, Ko2}.  
  
One can replace the $SU(2)_L$ doublets $(l_{LI},\,\! l_{RI})$  $(I=1,\,\! 2)$ with $SU(3)_c$ 
triplets, whose $U(1)_Y$ charges are $-1/3$ and $U(1)'$ charges are the same. One can also replace them with fields charged or not charged under $SU(3)_c \times SU(2)_L$. 
As discused in the section 4, we can also consider the case with $q_{i}=d_i=u_i=1/3$ and $ l_{i}=e_i=0$ where the anomaly condition for $U(1)_YU(1)'^2$ is satisfied without the SM gauge vector-like pairs.

We also have constraints on the $U(1)'$ charges from the Yukawa couplings 
which generate masses of the SM fermions and extra fermions.
Furthermore, our charge assignment may allow the mixing between the SM fermions 
and the extra fermions, which causes the flavor-changing neutral currents (FCNCs). 
For example, in the case that $q_h$, $l_i$ and $e_i$ are set to $0$,
we can generally write down the mixing terms between SM leptons and extra leptons, such as $m_i \overline{L'_R} L^i$.
We assume that such tree-level mixing terms are controlled to avoid the FCNC problem.

\section{Cold dark matter}
\label{sec:CDM}

As we discuss in the above, we require the extra fermions, depending on the charge assignment. 
In this section, we comment on the possibility that the required chiral fields 
give rise to CDM candidates, based on Ref. \cite{Ko1, Ko2}.

First, let us focus on models with $SU(2)_L$ doublet pairs, 
$(l_{LI},\,\! l_{RI})$.
$l_{LI}$ and $l_{RI}$  are chiral fields charged under $SU(2)_L \times U(1)_Y$
like the left-handed lepton,  so we can expect one component to be 
charged and the other to be neutral after EWSB.
The Yukawa couplings corresponding to the mass terms are given by the extra scalar, $\Phi$ with $U(1)'$ charge, $q_{\Phi}$.
When $q_{\Phi}$ is set to $1$, mass terms 
of$(l_{LI},\,\! l_{RI})$ (or ($q_{LI},\,\! q_{RI})$) are linear to the $\langle \Phi \rangle$, 
because the solution with $Q_L-Q_R=1$ is found in any cases.  

$l_{LI}$ and $l_{RI}$ do not have Yukawa couplings with the SM fermions 
because of $U(1)'$ symmetry, so that $U(1)$ global symmetry, which is phase rotation 
of $l_{LI}$ and $l_{RI}$, could be assigned, and their stabilities are 
guaranteed by the global symmetry. 
After EWSB, one component is charged and the other is neutral.
Their masses are degenerate before EWSB, but especially the charged particles
get enough large radiative corrections and then heavier than the neutral. 
If the mass split is bigger than the electron mass, the charged can decay 
to the neutral, $e$, and $\nu_e$ through the weak interaction.
We conclude that the light neutral components could be good CDM candidates.

The CDM can annihilate to 2 light SM fermions through $Z'$ and $Z $ boson 
exchanging, and 2 gauge boson in the $t$-channel, so that the thermal relic density
could be achieved by the gauge interactions.
The mass of the CDM is constrained by the search for extra leptons, 
that is, it should be heavier than $100$ GeV,
where the direct searches for dark matters shows negative results.
Even if we consider the scenario that $Z'$ interaction is negligible 
in the direct scattering,
$Z$ boson exchanging will work at the higher order. Such heavy CDM scenario 
is not favored by the direct searches, such as XENON100~\cite{Aprile:2011hi}.


Next, we discuss the case with extra $SU(3)_c$ triplets, 
$(q_{LI},\,\! q_{RI})$ or only the extra family, such as $Q_R'$, $U'_L$, and $D'_L$.
The mass terms of $(q_{LI},\,\! q_{RI})$ are give by $\Phi$ like the masses of $l_{LI}$ and $l_{RI}$.
The masses of the extra family are given by the $SU(2)_L$ doublet Higgs like the SM fermions.
Those extra colored particles would be stable because of the $U(1)'$ symmetry 
and the chirality.
In order to allow the decay of the extra colored particles, we introduce 
mixing term between the extra quarks and the SM quarks according to adding $X$,
where $X$ is a SM gauge-singlet scalar with $U(1)'$ charge.
$X$ has the Yukawa couplings, such as $ \lambda_i X^{\dagger} \overline{D_{Ri}} q_{L1}$,
and the extra quarks could decay to $X$ and the SM quarks if $X$ is lighter that the extra quarks.
If $X$ does not get nonzero vev, we can avoid tree-level FCNC and expect $X$ to be stable 
because of the remnant global symmetry after $U(1)'$ breaking. 
This type of dark matter has been well investigated in Ref.~\cite{Wise, Wise2, Alwall:2010jc},
and we discuss the dark matter physics in the gauged $U(1)_B (\equiv U(1)')$ model, in the next section.

\section{Supersymmetric Gauged $U(1)_B \times U(1)_L$ Model}
As an illustrative model, we discuss the supersymmetric gauged $U(1)_B \times U(1)_L$ model, 
which has been proposed in Ref.~\cite{Ko-Omura}. The gauged $U(1)_B$ is anomalous in the $SU(2)_L^2U(1)_B$ and the $U(1)_Y^2U(1)_B$ products, so that the extra generation, which was introduced in Sec.~2, is required.
Based on the argument in Sec.~3, the extra scalar, $X$, with nonzero $U(1)_B$ charge should be added and $X$ will
be a good CDM candidate \cite{Wise}. If we consider the extension to the supersymmetric, we can realize several CDM candidates, such as $X$, the superpartner of $X$, $\widetilde{X}$, and the neutralino, because of the remnant global $U(1)$ symmetry, which is originated by the gauged $U(1)_B$, and the R-parity, which could be realized spontaneously after $U(1)_B \times U(1)_L$ breaking \cite{Ko-Omura}. The CDMs, $X$ and $\widetilde{X}$, could annihilate to the SM particles through the $U(1)_B$ gauge boson ($Z_B$) exchanging, and the new gauge interaction could be strong, compared with the well-known gauge symmetry, such as $U(1)_{B-L}$. This is because such leptophobic charge assignment can evade the strong bound from the Drell-Yang process, assuming that the kinetic mixing is enough small. In the following, we discuss the dark matter physics and introduce the constraints on the leptophobic gauge interaction.

In this model, $X$ and $\widetilde{X}$ can annihilate to the SM quarks though the $Z_B$ exchanging.
If the Yukawa couplings with the extra quarks and the SM quarks are sizable, the CDMs can also annihilate 
in the $t$-channel. However, the Yukawa couplings should be controlled to avoid the FCNCs, so that
we simply assume that the Yukawa couplings are enough small, compared with the gauge interaction.
 
\begin{figure}
\includegraphics[width=0.45\textwidth]{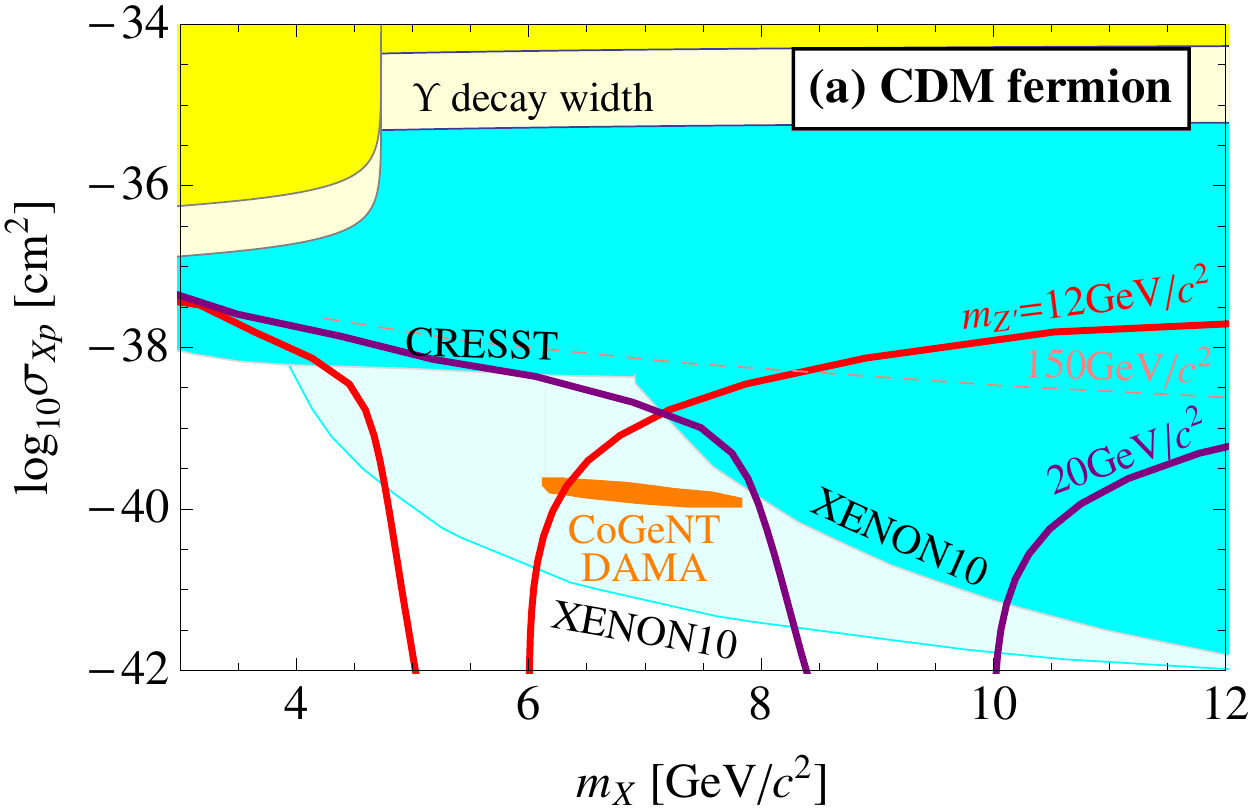}
\includegraphics[width=0.45\textwidth]{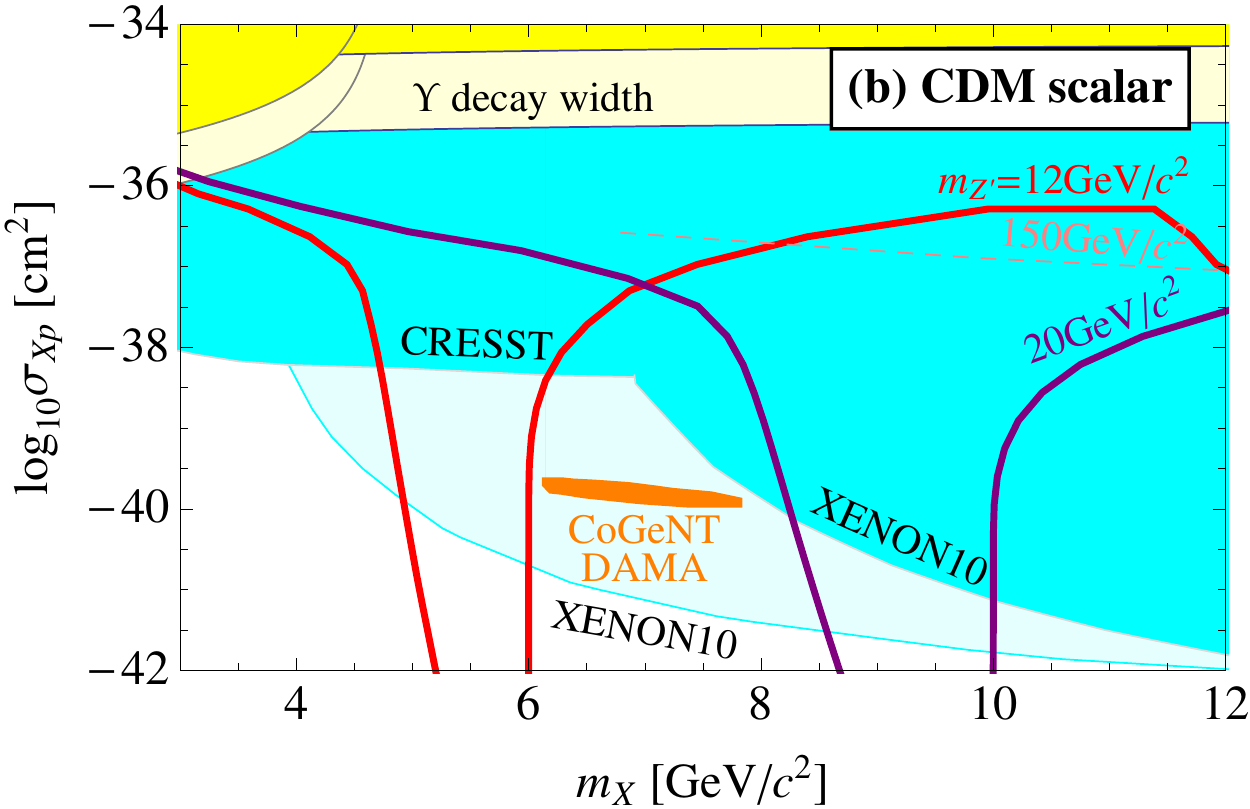}
\caption{Contour lines of $\Omega_Xh^2=0.1123$ for several values of the $Z_B$ boson mass $m_{Z_B}$~\cite{Gondolo}. On each contour, the cosmic density of particles $X$ (fermions in panel (a), and scalars in panel (b)) equals the cosmic density of CDM. Also shown are the DAMA/CoGeNT region (in orange), direct detection constraints (in blue)~\cite{XENON10}, and accelerator constraints (in yellow).}
\label{fig1}
\end{figure}

Depending on the mass spectrum of $X$ and $\widetilde{X}$, we can describe the each scenario:
$X$ or $\widetilde{X}$ is a CDM. The thermal relic density is given by the annihilation cross section of the CDM to the SM particles, and it is well known that the cross section should be $\langle \sigma v \rangle \sim 1 $pb to explain the WMAP data. $X$ and $\widetilde{X}$
annihilate to the SM quarks through the $Z_B$ exchanging: the cross section, $ \sigma v $, could be estimated as 
 $ \sigma v  \sim v^2 m_X^2 g_B^4/M^4_{Z_B}$ in the scalar case and $ \sigma v  \sim m_X^2 g_B^4/M^4_{Z_B}$ in the Dirac case, if the each CDM mass is far from the resonance, $m_X \sim M_{Z_B}/2$. On the other hand, the experiments of the direct detections succeed in limiting the scattering cross section of the CDM with nuclei. 
For example, if $m_X$ is $O(100)$GeV, the spin-independent cross section, $\sigma_{SI}$, should be smaller than $10^{-44}$cm$^2$~\cite{Aprile:2011hi}. Now $\sigma_{SI}$ is also linear to $g_B^4/M_{Z_B}^4$, so that $\langle \sigma v \rangle$ will be less than $1$pb, which induce the overproduction of the CDM. 
If $m_X$ is lower than the energy threshold of the direct detections, the upper bound on $g_B/M_{Z_B}$ would be relaxed. In Ref.~\cite{Gondolo, Ko3}, we discussed the scenario that $m_X \sim O(1)$GeV and $M_{Z_B} \sim 150$GeV, motivated by the $Wjj$ excess which was reported by the CDF collaboration~\cite{Aaltonen:2011mk}. 
In Fig.~\ref{fig1}, the dotted pink line also corresponds to the $M_{Z_B} = 150$GeV case with $\Omega_X h^2 =0.1123$. The CRESST experiment \cite{CRESST} gives the strongest bound and the bound is just below the dotted pink line.  
Furthermore, it has been pointed out that the mono-jet signals, $p \overline{p} \rightarrow g, Z_B \rightarrow g, X, X^\dagger (\widetilde{X}, \overline{\widetilde{X}})$, strongly constrain the heavy $Z_B$ scenario \cite{Ko3}. 
In fact, the bound seems to appear at the order, $\sigma_{SI} \sim 10^{-38}$cm$^2$ in Fig.~\ref{fig1}, which totally cover the dotted-pink line \cite{monojet}. 

On the other had, we could consider the scenario with very small $M_{Z_B}$ and small $g_B$, because of the leptophobic charge assignment. In this case, the mono-jet bound could be relaxed because of the small gauge coupling. As discussed in Ref. \cite{Ko-Omura, Gondolo},
the strongest bounds come from the invisible and hadronic decay widths of the $\Upsilon$ meson 
\cite{Carone:1995,Aranda:1998fr,Fayet:2009tv}. 
Contour lines of $\Omega_Xh^2=0.1123$ in the $m_X$--$\sigma_{Xp}$ plane are shown in Fig.~\ref{fig1} for light $Z_B$. The parameter $\alpha'$ changes along each line. Below each line, one has $\Omega_Xh^2>0.1123$. 
The thick red and purple contours correspond to $m_{Z_B}=$12 GeV$/c^2$ and 20 GeV$/c^2$, respectively. Each contour shows a dip at $m_X=m_{Z_B}/2$ due to the annihilation through the $Z_B$ resonance. 

Even in this case, the contour lines are just above the upper bound from the CRESST experiment, except for the resonance. On the other hand, around the resonance,
we see that the contour lines sweep the DAMA/CoGeNT region \cite{Hooper} for $Z_B$ masses in the range $\sim1$ to $\sim20$ GeV$/c^2$, touching the DAMA/COGeNT region on the left at the lowest $m_{Z_B}$ and on the right at highest $m_{Z_B}$.

\section{Summary}
We discussed $U(1)'$ models, especially motivated by CDM. Most of $U(1)'$ models, where the SM fermions are charged under the extra $U(1)'$ symmetry, are anomalous without extra chiral fermions. 
We introduced one extra generation with the opposite chirality, two SM vector-like pairs and scalars to achieve all anomaly conditions and to avoid conflicts with experiments. We found that the neutral particles,
which are included in the extra matters, become good CDM candidates because of the remnant global symmetry. 
As an illustrative model, we discussed the supersymmetric gauged $U(1)_B \times U(1)_L$ model, and investigated the relic density and the direct detection. Thanks to the leptophobic charge assignment, we can evade the strong bounds at LEP and Tevatron. If the $Z_B$ is heavy, the bound from the mono-jet signal may constrain the interaction of CDM strongly. We also discussed the light $Z_B$ scenario, and investigated how we can achieve the DAMA/CoGeNT signal. We have to require the fine tuning, $m_{Z_B} \sim 2 m_X$, but we can surely evade the bounds from collider experiments.      

\ack

I am grateful to P. Gondolo, P. Ko, and Chaehyun Yu for collaborations.

\end{document}